\documentclass[11pt]{article}
\usepackage{epsfig}
\usepackage{epsf}

\def \qqbar {q\bar q}
\def \bbbar {B\bar B}
\def \jp {J/\psi}
\def \llb {\Lambda\bar\Lambda}
\def \ssb {\Sigma^0\bar\Sigma^0}
\def \xxb {\Xi^-\bar\Xi^+}
\def \ee {e^+e^-}
\newcommand{\ang}[3] {\alpha'={4\alpha +(#1#2 \alpha )g_\sigma^4\over
4+#3(1+\alpha )g_\sigma^4}} 

\newcommand{\ppbar}{p\bar p}
\newcommand{\psip}{\psi(2S)}

\title{Rescattering effects of baryon and antibaryon  in heavy quarkonium decays}
\author{Hong Chen$^a$, Rong-Gang Ping$^b$\\
a) School of Physics, Southwest University, Chongqing, 400715, China. \\
b) Institute of High Energy Physics,Chinese Academy of Sciences,\\
 P.O. Box 918(4), Beijing 100049, China;\footnote{Mailing address}\footnote{E-mail:pingrg@mail.ihep.ac.cn}\\
\date{ }}
\begin{document}
\maketitle

\abstract{Rescattering effects of baryon and antibaryon in heavy
quarkonium decays are investigated by studying their angular
distributions. The rescattering amplitudes are phenomenologically
evaluated by modeling the intermediate range interaction as a
$\sigma$ or pion meson exchange between $\qqbar$ quarks. The results
show that the rescattering effects play an important role in
determination of the angular distribution in heavy quarkonium
decays. Especially, for $\jp$ and $\psip$ decays into $\llb$, $\ssb$
and $\xxb$ the angular distribution parameters could turn to be
negative values in the limit of helicity conservation. These results
provide us a possible explanation for understanding the negative
sign of the angular distribution parameter measured for
$\jp\to\Sigma^0\bar\Sigma^0$, namely, {\it it might come from the
baryonic SU(3)$_F$ symmetry breaking by incorporating
rescattering effects}.}\\
\vspace{0.0cm}
 \vspace{1.0cm}

\section{Introduction}
Exclusive decays of heavy quarkonium into baryon antibaryon pairs
$(B\bar B)$ are widely investigated as a laboratory to test
perturbative QCD (pQCD) properties or to study the baryonic
properties \cite{brodsky,kroll,carimalo, pingrga}. The branching
fractions of $\jp$ and $\psi'$ decays into $\bbbar$ have been
systematically studied in the framework of pQCD
\cite{brodsky,kroll}. The angular distributions of
$e^+e^-\to\jp,\psi'\to B_8\bar B_8$ ($B_8$: octet baryon), were also
investigated theoretically and experimentally by many groups
\cite{ang_theo_exp,plb591,plb632,prd29,plb610}. For $\jp$ produced
from $\ee$ annihilation, they must have spin $\pm1$ along the beam
direction. For exclusive decays of $\jp$ into octet baryonic pair,
the outgoing $\bbbar$ can be taken either S-wave or D-wave state in
laboratory system, thus the experimental observation of the outgoing
baryon direction takes the form $dN/d\cos\theta\propto
(1+\alpha\cos^2\theta)$, where $\theta$ is the angle between
outgoing baryon direction and $\ee$ beam direction. The angular
distribution parameters of $\alpha$ have been measured by many
groups \cite{plb632,prd29,plb610}. Recently, the BES collaboration
have released a more accurate measurement on the parameter $\alpha$
for $\jp\to\ppbar,~\llb$ and $\ssb$ decays with 58 million $\jp$
events accumulated at BEPC  as shown in Table \ref{expang}. It is
interesting to note that the parameter $\alpha$ for $\jp\to\ssb$ is
negative within $1\sigma$ accuracy, and the sign of the central
value of $\alpha$ for $\jp\to\xxb$ measured at Mark II is also
negative though its uncertainty is still very large.

\begin{table}[htbp]
\begin{center}
\parbox{0.7\textwidth}{\caption{Angular distribution parameter $\alpha$ for $\jp\to\bbbar$
decays. They are assumed to be the form of $dN/d\cos\theta\propto
1+\alpha\cos^2\theta$.\label{expang}}}
\begin{tabular}{lrcc}
\hline\hline
& & \multicolumn{2}{c}{Calculated value of $\alpha$}\\
Decay mode & Measured value of $\alpha$ & Ref. \cite{brodsky} & Ref.
\cite{carimalo}\\\hline
$\jp\to\ppbar$& $0.68\pm 0.06$\cite{plb591} &1 &0.69 \\
$\jp\to\llb$& $0.65\pm 0.11$\cite{plb632} & 1&0.51 \\
$\jp\to\ssb$& $-0.24\pm 0.20$\cite{plb632} &1 &0.43 \\
$\jp\to\xxb$& $-0.13\pm 0.59$\cite{prd29} &1 & 0.27\\
$\psi'\to\ppbar$&$0.67\pm0.16$\cite{plb610} &1 &0.80 \\
 \hline\hline
\end{tabular}
\end{center}
\end{table}
Theoretically, the study on the angular distribution parameter
provides us a copious information on polarization of decaying
particles , decay mechanisms, and strong interaction and so on.
Brodsky and Lepage once demonstrated that the exclusive process at
large momentum transfer $Q^2$ can be used to test the gluon spin and
other basic elements of perturbative quantum chromodynamics (QCD)
\cite{brodsky}. They verified  that the vector-gluon coupling
conserves quark helicity when quark and gluon masses are negligible
at the large scale $Q^2$. Since the hadronic helicity is the sum of
the helicities of its valence quarks, thus leads to the conclusion
that the total hadronic helicity is conserved up to corrections of
order $m_q/Q$ or higher. For $\ee\to\jp\to\bbbar$ decays, angular
distributions take the form of ${d\Gamma\over d\cos \theta}\propto
1+\cos^2\theta$ by the selection of the helicity conservation rule,
which gives the asymptotic value of $\alpha=1$. They conclude that
verifying the $1+\cos^2\theta$ angular distribution would provide
one with a clear proof of the spin one assumption for the gluon,
since with a scalar or a tensor gluon one would be led to a
$\sin^2\theta$ distribution.

In the past, this asymptotic value has been corrected by many
groups. In charmonium decays, we are far from an asymptotic regime
due to the factor that $m_B/m_\psi\sim 1/3$ is far from negligible.
The baryonic mass effects have already been investigated by
Claudson, Glashow and Wise \cite{cgw}, and they found a value of
$\alpha=0.46$ for $\jp\to\ppbar$, which confirms the idea of large
mass corrections to the asymptotic value. The quark mass effect has
been investigated by Carimalo \cite{carimalo}. It is assumed that
the quark created with effective mass $m$ inside a baryon carries
one-third of the total momentum of the out-going baryon, and the
three symmetric gluons carry the same momentum $(M_\psi/3)$ with no
other extra interaction. In this simple static-quark model, the
angular distribution can be explicitly expressed by
\begin{equation}\label{}
\alpha={(1+u)^2-u(1+6u)\over(1+u)^2+u(1+6u) } \textrm{~~with~~}
u={m_B^2\over m_\psi^2}.
\end{equation}
This formula gives the parameter $\alpha=0.69$ for $\jp\to\ppbar$,
very close to the experimental value. However, the calculated
branching fraction of $\jp\to\bbbar$ decays are inconsistent with
measured values with $1\sigma$ in this model. Some authors argued
that the non-perturbative effects of the baryon should be
parameterized in the calculation. Further investigations including
quark mass effects, electromagnetic effects and higher-twist
corrections to $\alpha$ have been made with a general conclusion of
$0<\alpha<1$ \cite{pingrga,ang_theo_exp}.

Phenomenologically, some authors argued that, in charmonium decays,
rescattering effects between the final particle states sometimes are
essential to explain experimental observations. For example, a
$\ppbar$ rescaterring near their threshold could make some
contributions to the observation on the near threshold-enhancement
of $\jp\to \gamma\ppbar$ \cite{fsi}. In meson sector, one example is
that the rescattering of $a_2\rho$ and $a_1\rho$ into $\rho\pi$ can
change the $\rho\pi$ production rate substantially, which could be
one of explanation for the $\rho\pi$ puzzle \cite{lbz}. Whether the
$\bbbar$ rescaterring could change their distribution significantly?
Stimulated by recent BES observations, we make a further
investigation on the angular distribution for heavy quarkonium
decays into $\bbbar$ by taking rescattering effects into account.

\section{Model and Formulation}
\begin{figure}[htbp]
 \vspace*{0cm}
\begin{center}
\epsfysize=4cm \epsffile{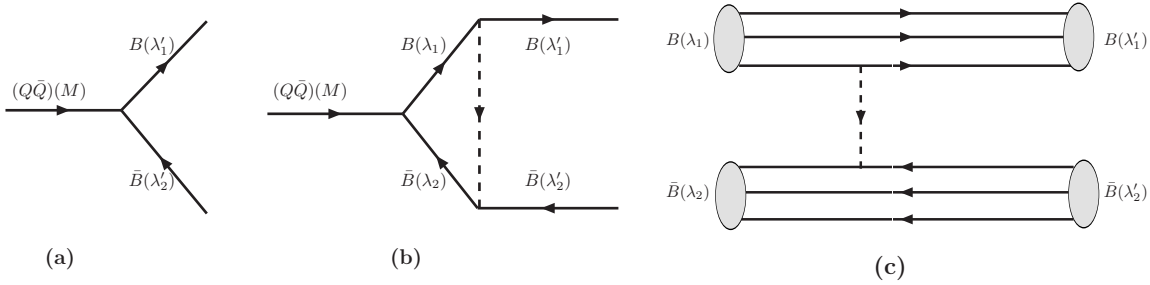}
\parbox{0.8\textwidth}{ \caption{The schematic illustration for heavy quarkonium decays into $\bbbar$ (a) and via
a rescattering process of $\bbbar$ (b), which is equivalent to a
exchange of a meson between $\qqbar$ quarks (c).}\label{fig1}}
\end{center}
\end{figure}

The $\bbbar$ produced from heavy quarkonium decays moves oppositely
with higher momentum (see Figure \ref{fig1}.(a)), they might
experience an intermediate-range interaction by exchange of mesons
(Figure \ref{fig1}.(b)). We assume that the outgoing baryonic pair
survives from the interaction without annihilation. We only consider
one-pion and tow-pion exchange processes. In principle, other mesons
with higher masses might be exchanged, their contributions are
supposed to be less than pion-exchange processes. As traditional
treatments to the $NN$ interaction, the process of two-pion exchange
is described by the picture of $\sigma$ meson
exchange\cite{machleidt}. For consideration of isospin conservation,
one-pion exchange could be allowed in $\ppbar$ and $\xxb$ decays
suppressed by a isospin factor of 1/3, and not allowed in $\llb$ and
$\ssb$ decays. As in chiral quark model\cite{huang}, we model the
intermediate range attraction in the $\bbbar$ interaction as a
rescattering of $\qqbar$ quarks by exchanging a $\sigma$ or pion
meson (Figure \ref{fig1}.(c)).

The helicity amplitude corresponding to Figure \ref{fig1} (a) and
(b) can be expressed as:
\begin{eqnarray}\label{}
M_1=D^{1*}_{M,\lambda_1'-\lambda_2'}(\theta_2,\phi_2)F_{\lambda_1',\lambda_2'},\textrm{~~~~for
Figure \ref{fig1} (a)}\\
M_2=\sum_{\lambda_1,\lambda_2,J}(2J+1)D^{1*}_{M,\lambda_1-\lambda_2}(\theta_1,\phi_1)F_{\lambda_1,\lambda_2}
T^J_{\lambda_1'\lambda_2',\lambda_1\lambda_2}D^{J*}_{\lambda_1-\lambda_2,\lambda_1'-\lambda_2'}(\theta_2,\phi_2),\textrm{~~~for
Figure \ref{fig1} (b)}
\end{eqnarray}
where $M,\lambda_1(\lambda_1')$ and $\lambda_2(\lambda_2')$ denote
the helicity of heavy quarkonium, baryon and antibaryon,
respectively. For heavy quarkonium production from positron
annihilation, $M=\pm1$. $F$ and $T$ are helicity amplitudes
corresponding to heavy quarkonium decays into $\bbbar$ and $\bbbar$
rescattering processes, respectively. For considering parity
invariance and time reversal invariance, one has $
F_{--}=F_{++},F_{-+}=F_{+-},
T_{-+--}=-T_{---+}=-T_{+-++}=T_{+++-}=T_{+---}=-T_{--++}=-T_{-+++}=T_{++-+}\equiv
T_1, T_{++--}=T_{--++}\equiv T_2,
T_{+-+-}=T_{-+-+}=T_{-++-}=T_{+--+}\equiv T_3$ and
$T_{++++}=T_{---}\equiv T_4$.

Note that interference terms vanishes between $M_1$ and $M_2$ due to
the orthogonal $D-$function, then the angular distribution of the
decayed baryon is expressed by:
\begin{equation}
{d\Gamma\over
d\textrm{cos}\theta_2}=\sum_{M,\lambda_1'\lambda_2'}\int
|M_1|^2d\phi_2+\int |M_2|^2 d\theta_1d\phi_1d\phi_2\propto 1+\alpha'
\textrm{cos}^2\theta_2,
\end{equation}
with
\begin{eqnarray}
 \alpha'&=&\{4\alpha-9\pi^2[(\alpha-1)(4T_2T_4+2T_2^2)+6T_1^2(\alpha+1)+\alpha(2T_4^2-3T_3^2)-2T_4^2-3T_3^2]\}\nonumber\\
    && /[4 +27\pi^2(2T_1^2+T_3^2)(1 + \alpha
    )],
\end{eqnarray}
where $\alpha={|F_{-+}|^2-2|F_{++}|^2\over |F_{-+}|^2+2|F_{++}|^2}$
is the net angular distribution parameter for $\bbbar$ without
rescattering effects. If the rescattering effects vanish, {\it i.e.
$T_i\to 0$}, then $\alpha'=\alpha$. It is worthwhile to note that
the sign of $\alpha'$ could be positive or negative determined by
$\bbbar$ scattering amplitudes.

Unfortunately, evaluation of rescattering amplitudes in the
framework of pQCD cannot be made without models, this is mainly due
to the fact that the non-perturbative information on bound states
can not be fully determined from a reliable method in theory. For
low lying states of octet baryons, we will employ the so-called
constituent quark model to evaluate rescattering amplitudes. For
$\sigma$ exchange, we define an operator
\begin{eqnarray}\label{}
A_{\lambda_1'\lambda_2',\lambda_1\lambda_2}(p_iq_ip_i'q_i',s_i \bar
s_is_i'\bar s_i')=\int {d^3x\over 2x^0}{1\over
x^2-m_\sigma^2+i\Gamma_\sigma m_\sigma}[\bar
u(p_3',s_3')g_\sigma u(p_3,s_3)\bar v(q_3,s_3)g_\sigma v(q_3+x,\bar s_3)\nonumber\\
\times
\delta^4(p_1-p_1')\delta^4(p_2-p_2')\delta^4(p_3-p_3'-x)+{\textrm{permutation
terms}}],
\end{eqnarray}
where $g_\sigma$ is the effective coupling strength of a $\sigma$
meson to $\qqbar$ quarks, which will be fixed by using the angular
distribution parameter of $\jp\to\ppbar$ as input value.
$\Gamma_\sigma$ and $m_\sigma$ are the $\sigma$'s decay width and
mass, respectively. $x$ denotes the transferred momentum.
$u(p_i,s_i)$ and $v(q_i,\bar s_i)$ are Dirac spinors for quarks and
antiquarks with helicity $s_i$ and $\bar s_i$, respectively. Their
normalization is taken as $\bar uu=-\bar vv=m/E$. For one-pion
exchange, the $g_\sigma$ should be replaced by $\gamma_5g_\pi$ and
propagator by $1/(x^2-m_\pi^2+i\epsilon)$. Then the rescattering
helicity amplitude is evaluated by:
\begin{eqnarray}\label{ha}
T_{\lambda_1'\lambda_2',\lambda_1\lambda_2}=\int \prod_{i=1,3}
d^3p_id^3q_id^3p_i'd^3q_i' \langle \psi_B(p_i',s_i')\psi_{\bar
B}(q_i',\bar s_i')|A_{\lambda_1'\lambda_2',\lambda_1\lambda_2}+A_{\lambda_1'\lambda_2',\lambda_2\lambda_1}\nonumber\\
+A_{\lambda_1\lambda_2,\lambda_1'\lambda_2'}+A_{\lambda_1\lambda_2,\lambda_2'\lambda_1'}|\psi_B(p_i,s_i)\psi_{\bar
B}( q_i,\bar s_i) \rangle\nonumber\\
\times \delta^3(\vec P-\vec p_1-\vec p_2-\vec p_3)\delta^3(\vec
P'-\vec p_1'-\vec p_2'-\vec p_3')\delta^3(\vec Q-\vec q_1-\vec
q_2-\vec q_3)\delta^3(\vec Q'-\vec q_1'-\vec q_2'-\vec q_3'),
\end{eqnarray}
where $\psi_B(p_i,s_i)\psi_{\bar B}(q_i,\bar s_i)$ is the product of
 $B$ and $\bar B$ wave function in momentum space, which includes the spin,
  flavor and spatial wave function. They are constructed in the naive quark
  model.
\section{Numerical Results}
\subsection{SU(6) basis}
For illuminating the role of the effect of $SU(3)_F$ symmetry
breaking in rescattering processes, we firstly evaluate rescattering
amplitudes by using SU(6) basis and then compare it with the
so-called $uds$ basis. We naively assume that the masses of quarks
satisfy $m_u=m_d=m_s=m_B/3$, where $m_B$ is the baryonic mass. This
means that we neglect binding energies in the baryonic bound state
and use the non-relativistic static quark model. The spin-flavor
wave functions of octet baryon are described by:
\begin{equation}\label{}
\psi_{SF}^B=\psi_{SF}^{\bar B}={1\over \sqrt
2}(\chi^\rho\phi^\rho+\chi^\lambda\phi^\lambda),
\end{equation}
where $\chi^\rho$ and $\chi^\lambda$ are the mixed-symmetry pair
spin-${1\over 2}$ wave function. For example, we have
\begin{eqnarray}\label{}
\chi^\rho_{1/2,1/2}&=&-{1\over \sqrt
6}\{|\uparrow\downarrow\uparrow\rangle+|\downarrow\uparrow\uparrow\rangle-2|\uparrow\uparrow\downarrow\rangle\}\nonumber
,\\
\chi^\lambda_{1/2,1/2}&=&{1\over \sqrt
2}\{|\uparrow\downarrow\uparrow\rangle-|\downarrow\uparrow\uparrow\rangle\},
\end{eqnarray}
for the case of the total spin ${1\over 2}$ and its projection
${1\over 2}$. The flavor wave function $\phi^\rho$ and
$\phi^\lambda$ are exactly analogous to the spin wave function but
in the flavor space. As in the most naive quark model, we assume
that the bound state wave function for constituent quarks can be
described by a simple harmonic-oscillator eigenfunction in their
center-of-mass (c.m.) system, i.e.
\begin{equation}\label{}
\phi(\vec k_\rho,\vec k_\lambda)={1\over (\pi\beta)^{3/2}}e^{-(\vec
k_\rho^2+\vec k_\lambda^2 )/2\beta}
\end{equation}
where $\beta$ is the harmonic-oscillator parameter and $\vec
k_\rho$, $\vec k_\lambda$ are defined as $ \vec k_\rho =(\vec
k_1+\vec k_2-2\vec k_3)/\sqrt 6$, and $\vec k_\lambda =(\vec
k_1-\vec k_2)/\sqrt 2 $ with $\vec k_1$, $\vec k_2$ and $\vec k_3$
the momenta for the three quarks in c.m. system of the corresponding
baryon.

\subsection{$uds$ basis}
To count for the SU(3)$_F$ symmetry breaking in strange baryons, we
use the so-called $uds$ basis, which makes explicit SU(3)$_F$
symmetry breaking under exchange of unequal mass quarks
\cite{capstick}. Flavor functions for strange baryons are taken as
\begin{eqnarray}
   \phi_\Lambda&=&{1\over \sqrt 2}(ud-du)s,  \\
   \phi_\Sigma&=&{1\over \sqrt 2}(ud+du)s,  \\
   \phi_\Xi&=&ssd,
\end{eqnarray}
and the construction of the spin wave function $\chi$ proceeds
exactly analogously to that of the flavor wave functions. For
instance, the spin wave functions for the total angular momentum
quantum 1/2 with their $z$ projection 1/2 are:
\begin{eqnarray}
   \chi_\Lambda&=&{1\over \sqrt 2}(|\uparrow\downarrow\uparrow\rangle-|\downarrow\uparrow\uparrow\rangle),  \\
   \chi_\Sigma&=&{1\over \sqrt 6}(|\uparrow\downarrow\uparrow\rangle+|\downarrow\uparrow\uparrow\rangle-2|\uparrow\uparrow\downarrow\rangle),  \\
   \chi_\Xi&=&{1\over \sqrt
   6}(|\uparrow\downarrow\uparrow\rangle+|\downarrow\uparrow\uparrow\rangle-2|\uparrow\uparrow\downarrow\rangle).
\end{eqnarray}
The spatial wave function for the ground state of the strange
baryons are chosen similarly to those of protons, but have an
asymmetry between the $\rho$ and $\lambda$ oscillators. In the
baryonic c.m. system, it reads:
\begin{equation}\label{}
\psi(\vec k_\rho,\vec k_\lambda)={1\over
(\pi^2\beta_\rho\beta_\lambda)^{3/4}}e^{-(\vec
k_\rho^2/2\beta_\rho+\vec k_\lambda^2/2\beta_\lambda)},
\end{equation}
where $\beta_\rho=(3km)^{1/2}$ and $m$ is the quark mass. For
example, we take $m=m_u$ for $\Lambda$, $\Sigma$, and take $m=m_s$
for $\Xi$. $\beta_\lambda=(3km)^{1/2}$ with
$m_\lambda=3mm_3/(2m+m_3)>m$ ($m_3=m_s$ for $\Lambda$, $\Sigma$, and
$m_3=m_u$ for $\Xi$). Using these equations, we relate
$\beta_\rho(\Lambda,\Sigma,\Xi)$ and
$\beta_\lambda(\Lambda,\Sigma,\Xi)$ to the harmonic-oscillator
parameter $\beta$ of nucleons, i.e.,
$\beta_\rho(\Lambda,\Sigma)=\beta,~\beta_\rho(\Xi)=\sqrt
{m_s/m_d}\beta$, and
$\beta_\lambda(\Lambda,\Sigma,\Xi)=\sqrt{m_\lambda/m}\beta_\rho(\Lambda,\Sigma,\Xi)$.
\begin{table}[htbp]
\begin{center}
\parbox{0.9\textwidth}{\caption{The angular distribution parameters for $\jp$, $\psi'$ and
$\Upsilon(1S)$ decays into $\ppbar$, $\llb$, $\ssb$ and $\xxb$
incorporated $\bbbar$ rescattering effects by exchange of a $\sigma$
meson. The parameters are set as $m_u=m_d=m_s=m_B/3$ and
$\beta=0.16\textrm{GeV}^2$ for SU(6) basis, while
$m_u=m_d=310\textrm{MeV},m_s=490\textrm{MeV}$ and
$\beta=0.16\textrm{GeV}^2$ for $ uds $ basis.\label{angf}}}
\begin{tabular}{rcc}
  \hline\hline
  Channel & SU(6) basis & $uds$ basis \\\hline
   $\jp\to\ppbar$& $\ang{14.8}{-8.5}{4.6}$ &$\ang{14.8}{-8.5}{4.6}$\\
   $\llb$&$\ang{14.8}{-8.3}{4.9}$&$\ang{13.1}{-164.9}{118.1}$\\
   $\ssb$&$\ang{13.5}{-7.6}{4.6}$&$\ang{11.6}{-25.7}{20.8}$\\
   $\xxb$&$\ang{8.7}{-4.9}{3.0}$&$\ang{7.0}{-15.4}{12.5}$\\\hline
   $\psi'\to\ppbar$&$\ang{14.8}{-8.9}{4.3}$&$\ang{14.8}{-8.9}{4.3}$\\
   $\llb$&$\ang{18.0}{-10.5}{5.5}$&$\ang{176.6}{-207.9}{139.1}$\\
   $\ssb$&$\ang{19.4}{-11.1}{6.0}$&$\ang{18.6}{-36.6}{30.2}$\\
   $\xxb$&$\ang{20.4}{-11.4}{6.6}$&$\ang{15.7}{-32.3}{26.5}$\\\hline
     $\Upsilon(1S)\to\ppbar$&$\ang{1.4}{-1.0}{0.3}$&$\ang{1.4}{-1.0}{0.3}$\\
   $\llb$&$\ang{2.1}{-1.5}{0.39}$&$\ang{25.6}{-26.8}{9.9}$\\
   $\ssb$&$\ang{2.4}{-1.7}{0.5}$&$\ang{2.7}{-4.5}{3.5}$\\
   $\xxb$&$\ang{3.1}{-2.1}{0.6}$&$\ang{1.7}{-3.4}{2.6}$\\
  \hline\hline
\end{tabular}
\end{center}
\end{table}

For numerical evaluation of rescattering amplitude of $\bbbar$,
there are three parameters to be determined, {\it i.e.} the quark
mass $m_u$, $m_s$ and the harmonic oscillator parameter $\beta$ in
nucleon spatial wave functions. As usually used in naive quark
model, we set $m_u=m_s=310$ MeV, $m_s=490$ MeV and
$\beta=0.16\textrm{~GeV}^2$. The rescattering amplitudes
$T_i(i=1,2,3,4)$ are obtained by evaluating the integration of
Equation (\ref{ha}) directly. The numerical program can be obtained
from authors. Then we list the angular distribution by incorporation
of $\bbbar$ rescattering effects in Table \ref{angf}.

The observed angular distribution of $\bbbar$ depends on the
effective coupling strength $g_\sigma$ and the net parameter
$\alpha$. In the chiral quark model, the coupling strength $g$ is
determined from $g_{\pi NN}$ \cite{huang}, which is $g=2.62$. Here
we fix this parameter by using the experimentally angular
distribution of $\jp\to\ppbar$ as an input value. For consistency
the net parameter $\alpha$ should be calculated from the same quark
model. Carimalo formula is a good approximation in the description
of the net parameter $\alpha$ by incorporating baryonic structure
information. If the quark mass is negligible or $4m_B^2\ll M^2$
($M:$ heavy quarkonium mass), the net parameter tends to the
asymptotic value $\alpha=1$. We will see how the observed angular
distribution parameter changes when the quark mass vanishes, so as
to check its asymptotic behavior under the condition of the helicity
conservation.

As shown in Table \ref{angn}, if the difference between the light
quark masses and strange quark masses is ignored and octet baryons
are described by SU(6) wave function, the observed angular
distribution parameters remain positive when the net parameter
$\alpha$ tends to the asymptotic value. Though the experimental
value for $\jp\to \llb$ seems to be consistent with results from
SU(6) basis, it is not enough to distinguish the SU(6) basis from
the $uds$ basis due to the uncertainty of the net parameter
$\alpha$. It is worthwhile to note that calculated values using the
$uds$ basis for $\jp$ decays into strange baryons change their signs
when the $\alpha$ tends to the asymptotic value. Though the
experimental error for $\jp\to\ssb$ is still larger, the measured
central value for this decay is negative within $1\sigma$ accuracy.
So this result provides us a possible explanation for understanding
the negative sign of angular distribution parameter for $\jp\to\ssb$
decay, namely, {\it it might come from the baryonic SU(3)$_F$
symmetry breaking by incorporating rescattering effects}.
 For further test of rescattering effects in charmonium decays, higher
accurate measurements of the angular distribution for
$\jp,\psi'\to\Sigma^0\bar\Sigma,\Xi^-\bar\Xi^+$ are desirable.

\begin{table}[htbp]
\begin{center}
\parbox{0.95\textwidth}{\caption{Comparison of the angular distribution parameter between
experimental and theoretical values by SU(6) basis and $uds$ basis.
Where the values of $\alpha'(\alpha=1)$ and $\alpha'$ (Carimalo)
indicate that the net parameter $\alpha$ are set to one and
evaluated by Carimalo formula, respectively. }\label{angn}}
\begin{tabular}{r|cc|ccc}
 \hline\hline
&\multicolumn{2}{c}{SU(6) basis}\vline& \multicolumn{2}{c}{$uds$
basis}&Experimental value
\\\cline{2-6}
 Channel
&$\alpha'(\alpha=1)$&$\alpha'$(Carimalo)&$\alpha'(\alpha=1)$&$\alpha'$(Carimalo)&$\alpha'$\\\hline
$\jp\to\ppbar$&0.68(input) &0.68(input)&0.68(input) &0.68(input)&$0.68\pm 0.06$ \cite{plb591}\\
$\Lambda\bar\Lambda$&0.66&0.69&-0.13&0.37&$0.65\pm0.11$ \cite{plb632}\\
$\Sigma^0\bar\Sigma^0$&0.65 &0.63&-0.34&0.23&$-0.24\pm0.20$ \cite{plb632}\\
$\Xi^-\bar\Xi^+$&0.64&0.45&-0.33&0.18 &$-0.13\pm0.59$
\cite{prd29}\\\hline
$\psi'\to\ppbar$&0.70 &0.84&0.70&0.84&$0.67\pm 0.16$ \cite{plb610}\\
$\Lambda\bar\Lambda$&0.69 &0.80&-0.11&0.20&...\\
$\Sigma^0\bar\Sigma^0$&0.68&0.78&-0.30&0.19&...\\
$\Xi^-\bar\Xi^+$&0.68&0.75&-0.31&-0.03&...\\\hline
$\Upsilon(1S)\to\ppbar$&0.84 &0.98&0.84&0.98&...\\
$\Lambda\bar\Lambda$&0.80&0.97&-0.06&0.59&...\\
$\Sigma^0\bar\Sigma^0$&0.80&0.96&-0.24&0.74&...\\
$\Xi^-\bar\Xi^+$&0.78&0.95&-0.31&-0.28&...\\\hline\hline
\end{tabular}
\end{center}
\end{table}

The case of one-pion exchange is also investigated in $\ppbar$ and
$\xxb$ decays under the SU(6) basis and $uds$ basis. Parameters are
chosen as same as the $\sigma$-exchange case. Numerical results show
that the contribution of one-pion exchange from the two
mixed-symmetry spin wave functions plays a different role in the
evaluation of rescattering amplitudes. Since the coupling of $\qqbar
\pi$ is selected as $g_\pi\gamma_5$, the contribution from the
$\lambda-$type spin wave function is dominant over $\rho-$type
functions. Thus the values of rescattering amplitudes for $\xxb$
decays under the $uds$ basis are quite smaller than those under the
SU(6) basis. With the same treatment as to the $\sigma$ exchange
case, the effective coupling strength of $g_\pi$ is determined by
using the experimental value of $\jp\to\ppbar$ decay as an input
value, we find out an approximate relation $\alpha'\simeq \alpha$
for $\jp$, $\psip$ and $\Upsilon(1S)$ decays into $\ppbar$ and
$\xxb$ channels under the two bases used.
\section{Discussion and Summary}
Based on our evaluation, we suggest that charmonium decays into
$\ssb$ and $\xxb$ could be used to look for $\bbbar$ rescattering
effects. For $\Upsilon(1S)\to\bbbar$ decays, the branching fraction
is relatively smaller than that for charmonium decays, furthermore,
its $\bbbar$ rescattering amplitudes are largely suppressed by a
factor of $1/M^2_{\Upsilon}$. It should be pointed out that the
$\bbbar$ angular distribution dose not become isotropic at the limit
of the baryonic zero-velocity, instead it is dominantly determined
by rescattering processes. From the point of this view, the
rescattering effects should be looked for in the decay with a slowly
moving $\bbbar$ pair.

It is instructive to compare our results with ones from the
quark-scalar-diquark model in \cite{kada}. In the asymptotic limit,
i.e., the helicity conservation, the sign of parameter $\alpha'$
from $uds$ basis turns to be negative. In \cite{kada} the negative
sign of this parameter appears when the masses of quarks, gluons and
baryons tend to zero for point-like baryons. This implies that if
some interactions between out-going quarks are considered, the sign
of the angular distribution parameter could flip in the asymptotic
limit of the helicity conservation.

To summarize: the rescattering effects of $\bbbar$ in heavy
quarkonium decays are investigated by studying the angular
distributions. The rescattering amplitudes are phenomenologically
evaluated by modeling the intermediate range interaction as a
$\sigma$ or pion meson exchange between $\qqbar$ quarks. The results
show that the rescattering effects play an important role in
determination of the angular distribution in heavy quarkonium decays
. To compare the calculated values by using the SU(6) basis and the
$uds$ basis, one finds that the parameter $\alpha'$ turns to be
negative values by suing the $uds$ basis in the limit of asymptotic
value $\alpha=1$. Even though the accurate values of $\alpha'$ are
not fully determined in our calculation due to uncertainties of the
net parameter $\alpha$, we conclude that these results still provide
us a possible explanation for understanding the negative sign of
angular distribution parameter measured in the decay
$\jp\to\Sigma^0\bar\Sigma^0$, namely, {\it it might come from the
baryonic SU(3)$_F$ symmetry breaking by incorporating rescattering
effects}.
 For further test of rescattering effects in charmonium decays, higher
accurate measurements of the angular distribution for
$\jp,\psi'\to\Sigma^0\bar\Sigma,\Xi^-\bar\Xi^+$ are desirable.

Acknowledgement: The work is partly supported by the National
Natural Science Foundation of China under Grant No.10575083, No.
10435080, No. 10375074 and No. 10491303.


\begin{thebibliography}{99}
\def\bescoa {{BES Collaboration, J.Z. Bai, { et.al.}}}
\def\bescob {{BES Collaboration, M.Ablikim, { et.al.}}}
\def\prd#1#2#3 {{~Phys. Rev. D {#1} (#3) #2 }}  
\def\plb#1#2#3 {{~Phys. Lett. B {#1} (#3) #2 }}  
\bibitem{brodsky}S. J. Brodsky and G. P. Lepage,
\prd{24}{2848}{1981} .
\bibitem{cgw}Mark Claudson, sheldon L. Glashow, and Mark B. Wise,
Phys. Rev. D 25 (1982) 1345.
\bibitem{kroll} Jan Bolz and Peter Kroll, Eur. Phys. J. C2 (1998) 545.
\bibitem{carimalo} C. Carimalo, Int. J. Mod. Phys. A2 (1987) 249.
\bibitem{pingrga} R. G. Ping, H. C. Chiang and B. S. Zou,
\prd{66}{054020}{2002} .
\bibitem{fsi}B. S. Zou and H. C. Chiang, Phys. Rev. D 69 (2004)
034004;\\
B. Kerbikov, A. Stavinsky, and V. Fedotov, Phys. Rev. D 69 (2004)
055205;\\
A. Sibirtsec, J. Haidenbauer, S. Krewald, Ulf-G. Mei$\beta$ner, and
A. W. Thomas, Phys. Rev. D71 (2005) 054010.
\bibitem{lbz} Xue-Qian Li, Divid V. Bugg and Bing-Song Zou, Phys.
Rev. D55 (1997) 1421.
\bibitem{ang_theo_exp}F. Murgia and M. Melis,
\prd{51}{3487}{1995} .
\bibitem{plb591}\bescoa, \plb{591}{42}{2004} .
\bibitem{plb632}\bescob, \plb{632}{181}{2006} .
\bibitem{prd29} M. W. Eaton, et.al. \prd{29}{804}{1984} .
\bibitem{plb610}Fermilab E835 Collaboration, A. Buzzo,
et.al.,\plb{610}{177}{2005} .
\bibitem{mark2xx} M. W. Eaton, et. al., \prd{29}{804}{1984} .
\bibitem{bes1xx} \bescoa, High Energ and Nuclear Physics, 26 (2002)
93.
\bibitem{machleidt} R. Machleidt, K. Holinde and Ch. Elster, Phys.
Rep. 149 (1987) 1.
\bibitem{huang} F. Huang and Z. Y. Zhang Phys. Rev. C 72 (2005)
024003.
\bibitem{capstick} S. Capstick and W. Roberts, Prog. Part. nucl.
Phys. {\bf 45} (2000) s241.
\bibitem{kada}E1-Hassan Kada and Joseph Parisi, Phys. Rev. D 47
(1993), 3967.
\end{thebibliography}
\end{document}